\documentclass[a4paper,11pt]{article}
\usepackage{pos}

\title{A quantum analytical Adam descent through parameter shift rule using {\tt Qibo}}

\author[a,b]{Matteo Robbiati}
\author[c]{Stavros Efthymiou}
\author[a,c]{Andrea Pasquale}
\author*[a,b,c]{Stefano Carrazza}

\affiliation[a]{TIF Lab, Dipartimento di Fisica, Universit\`a degli Studi di
	Milano and INFN Sezione di Milano\\ Via Celoria 16, 20133, Milan, Italy.}

\affiliation[b]{CERN, Theoretical Physics Department, CH-1211
	Geneva 23, Switzerland.}

\affiliation[c]{Quantum Research Centre, Technology Innovation Institute, Abu Dhabi, UAE.}

\emailAdd{matteo.robbiati@unimi.it}
\emailAdd{stavros.efthymiou@tii.ae}
\emailAdd{andrea.pasquale@unimi.it}
\emailAdd{stefano.carrazza@cern.ch}

\abstract{In this proceedings we present quantum machine learning optimization
	experiments using stochastic gradient descent with the parameter shift rule
	algorithm. We first describe the gradient evaluation algorithm and its
	optimization procedure implemented using the {\tt Qibo} framework. After
	numerically testing the implementation using quantum simulation on classical
	hardware, we perform successfully a full quantum hardware optimization
	exercise using a single superconducting qubit chip controlled by {\tt Qibo}.
	We show results for a quantum regression model by comparing simulation to
	real hardware optimization.}

\FullConference{%
	41st International Conference on High Energy physics - ICHEP2022\\
	6-13 July, 2022\\
	Bologna, Italy
}

\begin{document}
	\maketitle

	\section{Introduction}

	In this work we use {\tt Qibo}~\cite{qibo,Efthymiou_2022,qibo_acat21}, a
	full-stack open-source framework for quantum simulation, control and
	calibration, to perform a gradient-based optimization on a one qubit Quantum
	Process Unit (QPU).
	Specifically, we implement an Adam optimizer~\cite{Adam}, a stochastic gradient
	descent method. Classical machine learning
	strategies make use of the Back-Propagation algorithm~\cite{B-P} which
	requires to know the value of the target function in the middle of the
	propagation in order to estimate the errors made in the predictions.
	On the other hand, this cannot be done when the model is a Variational
        Quantum Circuit (VQC) because it would be necessary to perform a measurement in
    the middle of the propagation in order to know the value of the function,
	causing the system to collapse to one of the accessible states and the
	consequent loss of the information accumulated up to that moment.

        VQC requires a method for evaluating gradients which should be
	deployable on the quantum hardware available in the NISQ era~\cite{Preskill}. A method for
	evaluating gradients of a quantum circuit was proposed in 2018 and it is
	known as \textit{Parameter Shift Rule} (PSR) \cite{Mitarai, PSR}.
        In the next sections we first summarize the PSR algorithm and then we show
	results for a VQC regression optimization example where all results
	presented in this manuscript have been obtained using the latest backends
	available in {\tt Qibo} (v0.1.8): {\tt qibojit}~\cite{Efthymiou_2022}, for
	simulation; {\tt
	qibolab}\footnote{\url{https://github.com/qiboteam/qibolab}} for hardware
	control; and {\tt
	qibocal}\footnote{\url{https://github.com/qiboteam/qibocal}} for
	calibration, characterization and validation.

	\section{The Parameter Shift Rule}

	Let us consider a circuit $\mathcal{U}(\bm{\theta})$ which, when applied to
	the initial state $\ket{0}$, returns the final state
	$\mathcal{U}(\bm{\theta})\ket{0} = \ket{q_f}$. Let also $B$ be an observable
	we involve into the estimation of an output variable $y$\footnote{In this
	case we use $\braket{B}\equiv \text{Prob}(\ket{0}) - \text{Prob}(\ket{1})$
	as estimator of $y$.}. Finally, let us consider the unitary operator
	\begin{equation}
	\mathcal{G}=\exp \bigl[-i\mu G \bigr],
	\label{gate G}
	\end{equation}
	in which the variational parameter $\mu\in\bm{\theta}$ appears and which has
	 at most two eigenvalues $\pm r$. We are interested in evaluating
	 $\partial_{\mu}f$, where $f$ is defined as follows:
	\begin{equation}
	f(\bm{\theta}) \equiv \braket{0|\mathcal{U}^{\dag}(\bm{\theta})\, B \,\mathcal{U}(\bm{\theta})|0}.
	\end{equation}
	It can be shown that, if $\mu$ appears in a single gate in the form
	\eqref{gate G}, the target estimation can be written as follows:
	\begin{equation}
	\partial_{\mu}f = r \bigl[f(\mu^+) - f(\mu^-) \bigr],
	\label{psr}
	\end{equation}
	with $\mu^{\pm} = \mu \pm s$ and $s=\pi/4r$. We use the case presented in
	\cite{Mitarai}, which requires the Hermitian G picked from the set of
	rotation generators $\frac{1}{2}\{\sigma_x, \sigma_y, \sigma_z \}$. In this
	case we can use the remarkable values $s = \pi / 2$ and $r = 1/2$.

	\subsection{Evaluating gradients in a re-uploading strategy}

	The circuit we build up follows the idea presented in 2019 by Adrián
	Pérez-Salinas \textit{et al.} \cite{re-uploading}, also called as
	\textit{re-uploading strategy}. It allows to use a sequence of parametrized
	gates applied to a single qubit structure as model. Specifically, we
	summarize below the model used:
	\begin{center}
		\begin{quantikz}
			\lstick{$\ket{0}$} & \gate{\mathcal{U}_1(\bm{\theta}_1)} &  \push{ \,\,...\,\, }  & \gate{\mathcal{U}_{L} (\bm{\theta}_L)} & \qw & \meter{$\ket{0/1}$}
		\end{quantikz}
	\end{center}
	with an arbitrary number of layers $\mathcal{U}_k (\bm{\theta}_k)$, defined as follows:
	\begin{center}
		\begin{quantikz} \qw & \gate{\mathcal{U}_k (\bm{\theta}_k)} & \qw
		\end{quantikz}
		=
		\begin{quantikz} \qw & \gate{RY (\theta_{k,1} \cdot x + \theta_{k,2})} & \gate{RZ (\theta_{k,3})} & \qw
		\end{quantikz}.
	\end{center}

	Each layer involves two parametric rotation gates. In the first rotation, we
	impose that the angle is constructed by a combination of two variational
	parameters and $x$, which is an input data. By repeating this procedure
	multiple times, we obtain the effect of the data re-uploading strategy. As
	explained in the previous section, the PSR can be used for evaluating the
	derivative of a circuit with respect to a variational parameter which
	appears at the exponent in a gate like \eqref{gate G}. Considering our
	ansatz, when the feature is involved into the parameter's definition, the
	PSR takes into account the entire angle of rotation $\theta'$, which we
	calculate as $\theta' = \theta \cdot x$. Thus, we need to correct the
	formula dividing the shift parameter $s$ by $x$ when evaluating $f(\pm \mu)$
	and, once the two $f$'s are obtained, we must recombine the values in the
	following way:
	\begin{equation}
	\partial_{\mu}f = r \bigl[f(\mu^+) - f(\mu^-) \bigr] \cdot x.
	\end{equation}
	The effect of this correction can be seen in Figure~\ref{tf vs PSR with
	features}, where five coloured lines are shown, each of which corresponds to
	a randomly selected value of $\theta$ with which we initialised a circuit
	consisting of a rotation around the $X$ axis and for which the angle of
	rotation is calculated as $\theta' = \theta \cdot x$. We calculate the
	derivative with respect to $\theta$ of the term:
	\begin{equation}
	\hat{y} = \bigl(\braket{B}_x - c\bigr)^2,
	\end{equation}
	where $\braket{B}_x$ is obtained through the difference of the probabilities
	of occurrence of the states $\ket{0}$ and $\ket{1}$ after $N_{shots}$
	executions of the circuit and $c$ is a constant for which we choose the
	arbitrary value $c=0.2$. Finally, we plotted the differences obtained by
	calculating the derivatives with the PSR and the \texttt{GradientTape()}
	method of \texttt{TensorFlow}. We carry out this procedure with and without
	the correction outlined above.

	\begin{figure}
		\centering
		\includegraphics[width=0.9\linewidth]{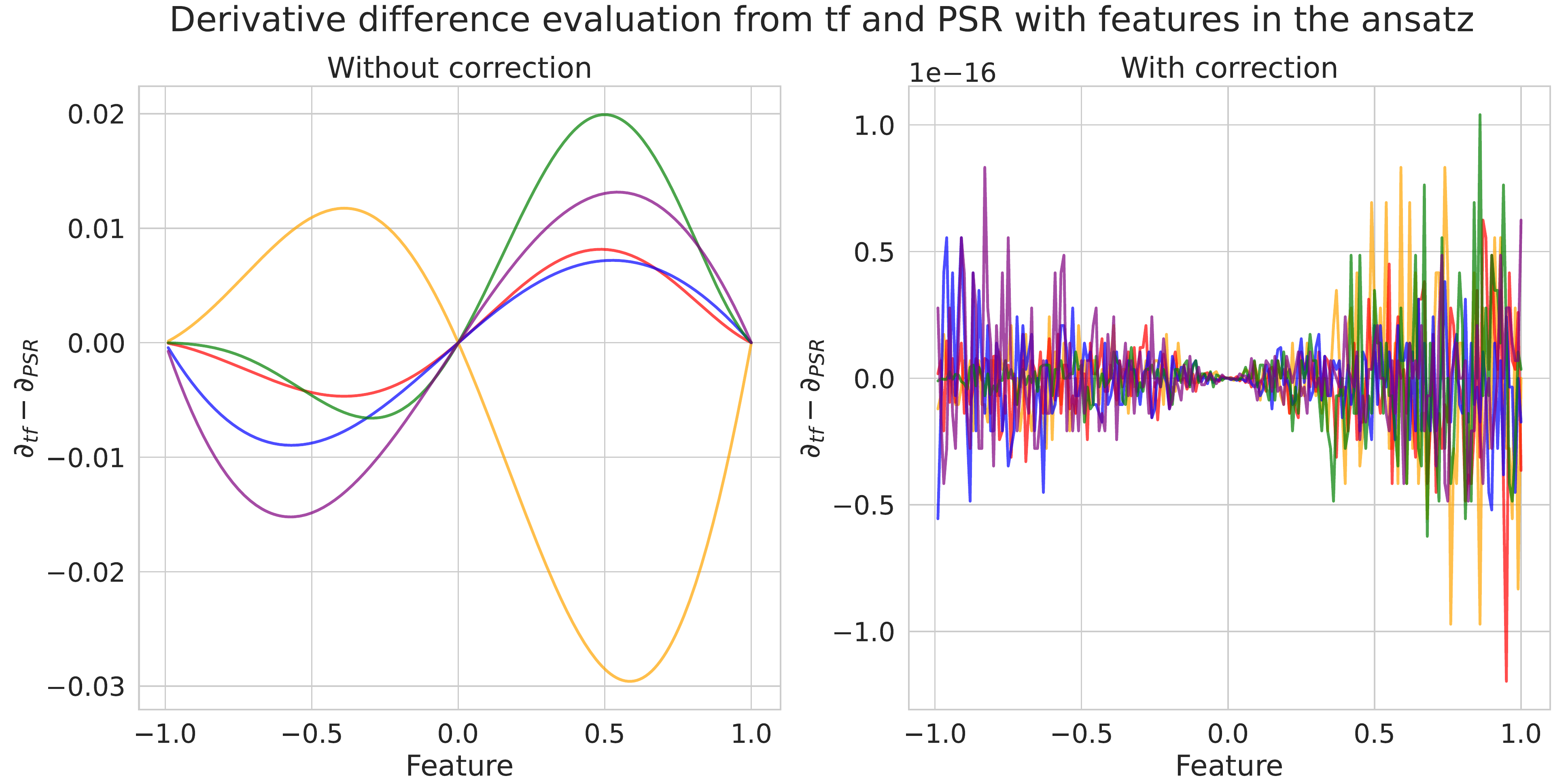}
		\caption{Difference between the derivative with respect to $\theta$ evaluated with the \texttt{TensorFlow}
		module \texttt{GradientTape()} called $\partial_{tf}$ and our PSR
		implementation called $\partial_{PSR}$. On the left, the value is
		calculated without making any change to the first version of the PSR. On
		the right, we can see the effect of the correction, which breaks down
		errors to the order of $10^{-16}$. Different colors correspond to
		different rotation's combinations.}
		\label{tf vs PSR with features}
	\end{figure}

	\subsection{Loss function's gradient}

	Our aim in this draft is to use PSR for gradient descent optimization. Since
	we choose a \textit{Mean Squared Error} loss function $J_{mse}$, we need two
	different estimators for calculating its derivative with respect to a $\mu$
	as explained in \cite{SGD_hybrid}. In fact, once we have chosen an input
	variable $x_j$ and defined the associated circuit via re-uploading strategy,
	we can execute it $N_{shots}$ times and calculate the following contribution
	to the loss function which we call $J_j$. At this point the derivative of
	$J_j$ with respect to $\mu$ becomes:
	\begin{equation}
	\partial_{\mu}J_j(\bm{\theta}) =
	2 \, \left (\braket{B}_{\bm{\theta}, x_j} -  y\right ) \,\partial_{\mu} \braket{B}_{\bm{\theta}, x_j},
	\label{dloss}
	\end{equation}
	where $\braket{B}_{\bm{\theta},x_j}$ is the estimator of $y$, calculated as difference of the probabilities of occurrence of the
	two fundamental states after performing the measurements on $\ket{q_f}$ and $\partial_{\mu} \braket{B}_{\bm{\theta}, x_j}$
	can be evaluated using the PSR. With the subscript $x_j$ we are emphasising that the circuit under consideration has a structure
	that depends explicitly on the variable $x_j$. This procedure must be carried out for each of the $p$ variational parameters
	involved in the definition of the model.
	Of course, $N_{data}$ can be used for the training, involving a computational cost equal to $\mathcal{O}(3 p N_{data}N_{shots})$
	for each optimization step. This aspect makes PSR a very computationally heavy technique.
	Nevertheless, it remains one of the few possible solutions for successfully performing gradient descent on quantum hardware.
	Before moving on to the description of the results obtained, let us summarise the algorithm we use for evaluating the derivative of
	$J_{mse}$ in the following block of pseudo-code.
	\begin{tcolorbox}
		\textbf{Optimization with parameter-shift rule}
		\begin{pseudo}[kw]
			\tn{initialize}  $ \partial J_{mse} = \vec{0}$   \\
			for \tn{\code{this\_feature}  and \code{this\_label}} in \tn{range $N_{data}$}: \\+
			\tn{define \code{this\_circuit(this\_feature)}}\\

			\tn{estimate $\braket{B}$ with $N_{shots}$ of \code{this\_circuit}} \\
			for \tn{$\mu$} in range \tn{$p$}: \\+
			\tn{with \eqref{psr} evaluate $\partial_{\mu} \braket{B}$ } \\
			\tn{use $\braket{B}$ and \eqref{psr} for calculate  : \code{this\_dloss} through  \eqref{dloss}}\\
			\tn{$ \partial _\mu J_{mse}$ += \code{this\_dloss}.}
		\end{pseudo}
	\end{tcolorbox}

	\normalsize
	\section{Deployment on quantum hardware}

	We can execute the code implementation presented above directly on the QPU
	by choosing the appropriate {\tt Qibo} backend. In this way we can perform
	the fit of $y=\sin 2 x$ with $x\in[-1,1]$ directly on quantum hardware. For
	this exercise we use the single qubit platform located in the Quantum
	Research Centre (QRC) of the Technology Innovation Insitute (TII) in Abu
	Dhabi\footnote{\url{https://www.tii.ae/quantum}}. The selected optimization
	hyper-parameters are reported in Table~\ref{hyperparameters} where $\eta$
	and $\varepsilon_J$ are respectively the Adam's learning rate and a
	threshold value for $J$ we impose for stopping the optimization.

	\begin{table}
		\centering
		\begin{tabular}{cccccc}
			\hline \hline
			$N_{\rm data}$  &
			$N_{\rm shots}$ &
			$\eta$ &
			$N_{\rm epochs}$  & $\varepsilon_{J}$\\
			\hline
			$25$ & $1024$ & $0.1$ & $100$ & $5\cdot 10^{-3}$ \\
			\hline \hline
		\end{tabular}
                \caption{Hyper-parameters used to train the qubit.}
		\label{hyperparameters}
	\end{table}

\begin{figure}
		\centering
		\includegraphics[width=0.9\linewidth]{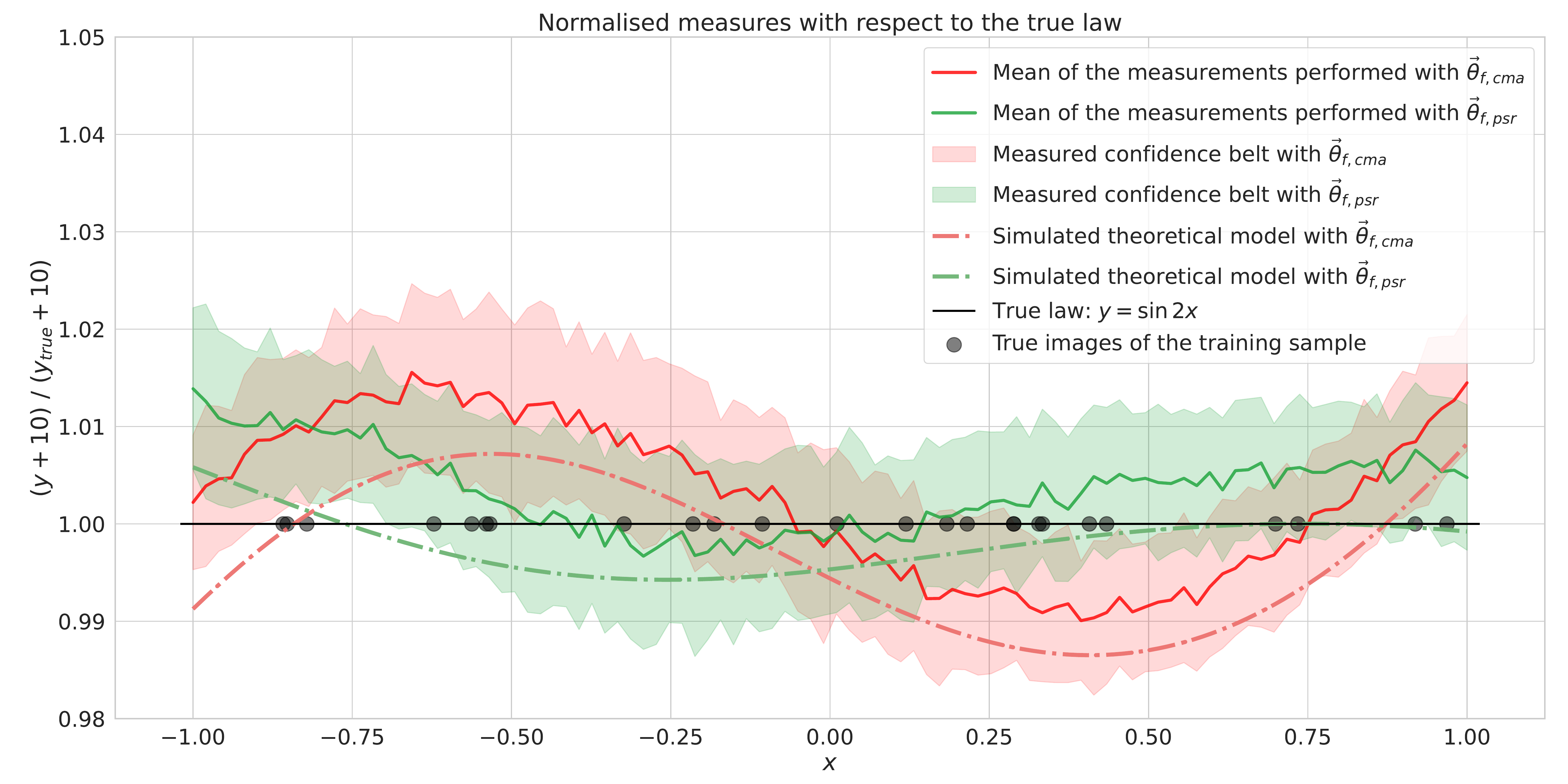}
		\caption{Results obtained by performing the optimization on the QPU and normalized with respect to the true law.
		The choice of adding $k=10$ to each result is entirely arbitrary and serves to prevent the values under consideration
		from diverging for $x$ tending to zero.
		The black line is the true law $\sin 2x$ and the black points are the true images of the training set.
		Colors green and red respectively refers to our PSR and to the CMA optimization.
		The dashed lines are the prediction's sets purposed through simulated theoretical models.
		The continuum lines and the confidence belts are drawn using one hundred predictions for each of one hundred points
		equally distributed into the interval $[-1,1]$.}
		\label{results}
	\end{figure}

	At first, we fix the initial parameters $\bm{\theta}_0$ and the training
	set. At this point we perform the optimization on the qubit using {\tt
	qibolab}, stopping it once reached the $\varepsilon_J$ value, obtaining the
	$\bm{\theta}_{best}$. We use the trained model to make statistics about the
	predictions; we pick one hundred points equally distributed in the $[-1, 1]$
        range and for each point $x_j$ we evaluate one hundred
	times the prediction. Finally, we calculate mean and standard deviation from
	the mean for each $j$. We use them for drawing the predictions as continuous
	line and the confidence bands in Figure~\ref{results}.
        Secondly, we use $\bm{\theta}_{best}$ obtained above for simulating a
	quantum circuit on a classical hardware and for getting the
	\textit{simulated theoretical} predictions. This passage is crucial to
	highlight the difference between the noisy quantum hardware and the
	classical simulation. The simulated predictions are presented as dashed
	lines in Figure~\ref{results}.
	To give consistency to the analysis, we compare the results obtained using
	this optimizer with those obtained using a genetic algorithm (CMA-ES). We
	also decide to present the results normalized with respect to the true law,
	which is represented as black line in Figure \ref{results}.
	We first note that the simulated theoretical model is compatible with the
	measurements recorded by the qubit, within the compatibility range we have
        defined. This is the case for both optimization methods. Furthermore,
	looking at the theoretical law, we see that it falls within the confidence
	belt relative to our training for a good part of the domain. The model
	proposed by the qubit, in this range, is compatible with the target law: we
        have successfully performed a gradient descent on a QPU with one qubit.

	We conclude the discussion by highlighting that {\tt Qibo} has achieved the
	level of completion required to obtain a successful application through the
	open-source availability of modules and backends for simulation, control and
	calibration. 

        \paragraph*{Acknowledgements} MR and SC are supported by CERN QTI program. SC
	is supported by the European Research Council under the European Union's
	Horizon 2020 research and innovation Programme (grant agreement number
	740006). The {\tt Qibo} project is supported by QRC.

	\bibliographystyle{unsrt}
	{\small
	\bibliography{references}

\begin{thebibliography}{10}

\bibitem{qibo}
Stavros Efthymiou, Sergi Ramos-Calderer, Carlos Bravo-Prieto, Adri\'an
  P{\'{e}}rez-Salinas, Diego Garc{\'{\i}}a-Mart{\'{\i}}n, Artur Garcia-Saez,
  Jos{\'{e}}~Ignacio Latorre, and Stefano Carrazza.
\newblock Qibo: a framework for quantum simulation with hardware acceleration.
\newblock {\em Quantum Science and Technology}, 7(1):015018, dec 2021.

\bibitem{Efthymiou_2022}
Stavros Efthymiou, Marco Lazzarin, Andrea Pasquale, and Stefano Carrazza.
\newblock Quantum simulation with just-in-time compilation.
\newblock {\em Quantum}, 6:814, sep 2022.

\bibitem{qibo_acat21}
Stefano Carrazza, Stavros Efthymiou, Marco Lazzarin, and Andrea Pasquale.
\newblock An open-source modular framework for quantum computing, 2022.

\bibitem{Adam}
Diederik~P. Kingma and Jimmy Ba.
\newblock Adam: A method for stochastic optimization, 2014.

\bibitem{B-P}
Ronald J.~Williams David E.~Rumelhart, Geoffrey E.~Hinton.
\newblock Learning representations by back-propagating errors.
\newblock {\em Nature}, 1986.

\bibitem{Preskill}
John Preskill.
\newblock Quantum computing in the {NISQ} era and beyond.
\newblock {\em Quantum}, 2:79, aug 2018.

\bibitem{Mitarai}
K.~Mitarai, M.~Negoro, M.~Kitagawa, and K.~Fujii.
\newblock Quantum circuit learning.
\newblock {\em Physical Review A}, 98(3), sep 2018.

\bibitem{PSR}
Maria Schuld, Ville Bergholm, Christian Gogolin, Josh Izaac, and Nathan
  Killoran.
\newblock Evaluating analytic gradients on quantum hardware.
\newblock {\em Physical Review A}, 99(3), mar 2019.

\bibitem{re-uploading}
Adri\'an P{\'{e}}rez-Salinas, Alba Cervera-Lierta, Elies Gil-Fuster, and
  Jos{\'{e}}~I. Latorre.
\newblock Data re-uploading for a universal quantum classifier.
\newblock {\em Quantum}, 4:226, feb 2020.

\bibitem{SGD_hybrid}
Ryan Sweke, Frederik Wilde, Johannes Meyer, Maria Schuld, Paul~K. Faehrmann,
  Barth{\'{e} }l{\'{e}}my Meynard-Piganeau, and Jens Eisert.
\newblock Stochastic gradient descent for hybrid quantum-classical
  optimization.
\newblock {\em Quantum}, 4:314, aug 2020.

\end{thebibliography}
	}
\end{document}